\documentclass[12pt,twoside]{article}
\catcode`\@=11

\def\newsymbol#1#2#3#4#5{\let\next@\relax%
 \ifnum#2=\@nehse%
 \ifnum#2=\tw@\let\next@\msyfam@\fi\fi%
 \mathchardef#1="#3\next@#4#5}
\def\mathhexbox@#1#2#3{\relax%
 \ifmmode\mathpalette{}{\m@th\mnnathchar"#1#2#3}
 hse\leavevmode\hbox{$\m@th\mathchar"#1#2#3$}\fi}
\font\tenmsy=msbm10
\font\sevenmsy=msbm7
\font\fivemsy=msbm5
\newfam\msyfam
\textfont\msyfam=\tenmsy
\scriptfont\msyfam=\sevenmsy
\scriptscriptfont\msyfam=\fivemsy
\edef\msyfam@{\hexnumber@\msyfam}
\def\Bbb#1{\fam\msyfam\relax#1}
\catcode`\@=\active

\newtheorem{theorem}{Theorem}[section]
\newtheorem{proposition}[theorem]{Proposition}
\newtheorem{lemma}[theorem]{Lemma}
\newtheorem{corollary}[theorem]{Corollary}

\newtheorem{remark}[theorem]{Remark}

\newcommand{\N}{{C_N}}
\newcommand{\WWW}{W}
\newcommand{\eq}[1]{\begin{equation}\label{#1}}
\newcommand{\en}{\end{equation}}

\newcommand{\proof}{{\noindent \it Proof:\ }}
\newcommand{\qed}{\hfill ${\rm QED}$\par\medskip}
\newcommand{\BR}{{{\Bbb R}^d}}
\newcommand{\bi}{\begin{description}}
\newcommand{\ei}{\end{description} }

\newcommand{\RR}{{\Bbb R}}
\newcommand{\hp}{H_{\rm p}} 
\newcommand{\eff}{{V_{\rm eff}}}
\newcommand{\heff}{H_{\rm eff}}
\newcommand{\mass}{m_{\rm eff}}
\newcommand{\e}{\epsilon}
\newcommand{\bl}[1]{\begin{lemma}\label{#1}}
\newcommand{\el}{\end{lemma}}
\newcommand{\bc}[1]{\begin{corollary}\label{#1}}
\newcommand{\ec}{\end{corollary}}
\newcommand{\bt}[1]{\begin{theorem}\label{#1}}
\newcommand{\et}{\end{theorem}}
\newcommand{\bp}[1]{\begin{proposition}\label{#1}}
\newcommand{\ep}{\end{proposition}}
\newcommand{\br}[1]{\begin{remark}\label{#1}}
\newcommand{\er}{\end{remark}}

\newcommand{\kak}[1]{(\ref{#1})}
\newcommand{\LR}{{L^2(\BR)}}

\newcommand{\fff}{{\cal F}}

\newcommand{\is}{\inf\!\sigma}

\newcommand{\lk}{\left(}
\newcommand{\rk}{\right)}
\newcommand{\lkk}{\left\{}
\newcommand{\rkk}{\right\}}

\newcommand{\add}{a^{\ast}}

\newcommand{\ov}[1]{\overline{#1}}
\newcommand{\hf}{H_{\rm f}}

\newcommand{\half}{\frac{1}{2}}

\newcommand{\hi}{H_{\rm I}}

\newcommand{\rh}{\hat \rho}
\newcommand{\la}{\hat \lambda}
\newcommand{\s}{\sigma}
\newcommand{\hhh}{{\cal H}}

\renewcommand{\d}{\displaystyle}
\newcommand{\non}{\nonumber}

\usepackage{amsmath,amssymb}

\newcommand{\essinf}{\operatorname*{ess.inf}}

\newcommand{\cE}{\mathcal{E}}

\newcommand{\slim}{\mathop{\mbox{\rm s-lim}}}

\newcommand{\supp}{\mathop{\mathrm{supp}}}
\usepackage{graphicx,color}

\makeatletter
\@addtoreset{equation}{section}
\makeatother

\title
{Enhanced binding for $N$-particle system interacting with 
a scalar bose field I}

\author{
Fumio Hiroshima\thanks{
Faculty of Mathematics, 
Kyushu University,   Fukuoka 812-8581, Japan.}
 and Itaru Sasaki\thanks{Department of Mathematics, Faculty of Science, 
Hokkaido University. }
}\date{\today}
\begin{document}
\pagestyle{myheadings}
\markboth{Enhanced binding}{Enhanced binding}

\setlength{\baselineskip}{15pt}
\maketitle

\begin{abstract}
An enhanced binding of an $N$-particle system 
interacting through a scalar 
bose field is investigated, where {$N\geq 2$}.  
It is not assumed that  this system has 
a  ground state for a zero coupling. 
It is shown, however,  that there exists a ground state  
for a sufficiently large values of coupling constants. 
When  the coupling constant is sufficiently 
large, $N$ particles are bound to each other 
by the scalar bose field, and 
are trapped by  external potentials.
Basic ideas {of the proofs} in this paper are applications of 
a weak coupling limit and a modified HVZ theorem. 
\end{abstract}
{\footnotesize
\tableofcontents
}
\section{Introduction}
In this paper we are concerned with an  enhanced binding of 
an $N$-particle system 
interacting with a scalar bose field. 
Here we assume that $N\geq 2$ and impose ultraviolet cutoffs on the 
scalar bose field. 
It may be expected that when $N$ particles 
interact with each other through a scalar bose field, 
a strong coupling enhances  the binding of this system 
if  forces mediating 
between each two particles are attractive. 
We want to justify this heuristic consideration for 
a certain quantum field model, which is  so-called 
the  Nelson model \cite{ne}. 
\subsection{The Nelson model}
We begin with giving the definition of the Nelson model. 
In this paper we denote the scalar product and the norm on 
a Hilbert space ${\cal K}$ by 
$(f,g)_{\cal K}$ and $\|f\|_{\cal K}$, respectively. 
Here  $(f,g)_{\cal K}$ is linear in $g$ and antilinear in $f$. 
Unless confusions arise, we omit the suffix ${\cal K}$.  
Let $\fff$ be the Boson Fock space over $\LR$ defined by 
$\fff:=\bigoplus_{n=0}^\infty [
\otimes_{s}^n \LR]$, 
where $\otimes_s^n\LR$ denotes the $n$-fold symmetric tensor product of 
$\LR$ with $\otimes_s^0\LR:={\Bbb C}$.
Vector $\Psi\in \mathcal{F}$ is written as 
$\Psi=\{\Psi^{(n)}\}_{n=0}^\infty$ 
with $\Psi^{(n)}\in \otimes_s^n L^2(\BR)$.
The Fock vacuum $\Omega\in\fff$ is defined by 
$\Omega:=\{1,0,0,\ldots\}$.  
$a(f)$ and $\add(f)$, $f\in\LR$,  denote the annihilation operator 
and the creation operator in $\fff$, 
respectively, which are defined by 
\begin{align*}
  (a(f)\Psi)^{(n)}(k_1,\ldots,k_n) &:= \sqrt{n+1}\int_{\BR} f(k)\Psi^{(n+1)}
(k,k_1,\ldots,k_n)dk,\quad n=0,1,2,\ldots
\end{align*}
with 
$$D(a(f)):=\Big\{\Psi\in \fff\Big| \sum_{n=0}^\infty \|(a(f)\Psi)^{(n)}
\|^2_{\otimes^n_s\LR}<\infty\Big\}
$$
and 
$\add(f):=(a(\bar f))^\ast$. They 
satisfy canonical commutation relations:  
$$[a(f), \add(g)]=(\bar f, g),\ \ \ [a(f), a(g)]=0=[\add(f),\add(g)],$$
on 
$
\mathcal{F}_0 := \{\Psi\in \fff| \Psi^{(n)}=0 \text{ for all }
  n\geq n_0 \text{ with some } n_0\}$. 
We informally write  as 
$$a(f)=\int_\BR a(k) f(k) dk,\quad 
\add(f)=\int_\BR \add(k) f(k) dk.$$
The Nelson Hamiltonian $H$ is a self-adjoint operator 
acting on  the  Hilbert space 
$$\hhh:=L^2(\RR^{dN})\otimes\fff,$$ 
which is defined by 
\begin{eqnarray*}
&& 
H:=H_0+\hi,\\
&&
H_0:=\hp\otimes 1+1\otimes \hf.
\end{eqnarray*}
Here $\hp$ is the $N$-particle Hamiltonian 
defined by  
$$
\hp:=\sum_{j=1}^N  \left( -\frac{1}{2m_j}\Delta_j+V_j \right),$$
where 
$m_j$  is the mass of the $j$-th particle. 
$\hf$ is the free Hamiltonian of $\fff$ given by 
$$\hf:=\bigoplus_{n=0}^\infty\left(
\sum_{j=1}^n \underbrace{
1\otimes\cdots\otimes \stackrel{j\text{-th}}{\check{\omega}}\otimes\cdots\otimes 1}_{n}
\right)
$$
with the dispersion relation $\omega(k):=|k|$, 
which is informally written as 
$$\hf=\int_\BR\omega(k)\add(k) a(k) dk.$$
Note that 
\begin{align*}
 (\hf\Psi)^{(n)}(k_1,\ldots,k_n)&=
(\omega(k_1)+\cdots+\omega(k_n))\Psi^{(n)}(k_1,\ldots,k_n),
  \quad n\geq 1, \\
  \hf\Omega&=0.
\end{align*}
It is well known that $\s(\hf)=[0,\infty)$,  $\s_{\rm p}(\hf)=\{0\}$, 
where $\s(K)$ (resp. $\s_{\rm p}(K)$) denotes the spectrum (resp. point spectrum) 
of $K$. Notation $\s_{\rm ess}(K)$ (resp. $\s_{\rm disc}(K)$)  
denotes  the essential spectrum (resp. discrete spectrum)  of $K$. 
We identify as
$$\hhh \cong \int^\oplus_{\RR^{dN}} \fff dx,$$
where $\int_{\RR^{dN}}^\oplus \cdots dx$ denotes  
a constant fiber direct integral \cite{rs4}
and $x=(x_1,\ldots,x_N)\in \mathbb{R}^{dN}$  
the position of  particles.
Finally $\hi$ denotes the interaction between $N$ particles and the scalar field given by 
$$
\hi:=\sum_{j=1}^N 
\alpha_j\int^\oplus_{\RR^{dN}} \phi_j(x_j) dx, 
$$
where $\alpha_j$'s  are real coupling constants and the scalar field 
$\phi_j(x)$ is given by 
$$\phi_j(x):=\frac{1}{\sqrt 2}\int_{\mathbb{R}^d}   
(\add(k) \hat \lambda_j(-k) e^{-ikx}+a(k) \hat \lambda_j(k) e^{ikx} ) dk,$$
where 
$\la_j$'s are ultraviolet cutoff functions.  
Note that 
$$(\hi\Psi)(x)=\sum_{j=1}^N \alpha_j\phi_j(x) \Psi(x),\quad a.e.x\in\RR^{dN},$$
with 
$$D(\hi):=\Big\{\Psi\in \hhh\Big| \Psi(x)\in \cap_{j=1}^N D(\phi_j(x))\mbox{ and } 
\sum_{j=1}^ N\int_{\RR^{dN}}
\|\phi_j(x)\Psi(x)\|_\fff^2 dx<\infty\Big\}.$$
Ground states of $H$ are defined by eigenvectors associated with 
eigenvalue $\is (H)$. 
We want to show the existence of ground states of $H$.
Generally $\is(H)$ is the bottom  
of  the essential spectrum of $H$.   
Although this makes troublesome to show the existence of ground states, 
it has been shown for various  models in quantum field theory by many authors, e.g., \cite{gll}, 
where one fundamental assumption is that 
$\hp$ has a ground state. 
In this paper we do not assume the existence 
of ground states of  $\hp$, 
which implies the absence of ground state of $H$ 
with $\alpha_1=\cdots=\alpha_N=0$, 
and 
show that $H$ has a ground state for sufficiently large values of 
coupling constants. 
This phenomena, if it exists, is called the enhanced binding.

\subsection{Weak coupling limits}
In our model under consideration, it is seen that 
the enhanced binding is derived  from the effective potential $\eff$ which is the sum of potentials  
between two particles. The effective potential can be derived from 
a {\it weak coupling limit} \cite{da,da2,h5,h6}, 
which is one of a key ingredient of this paper.
Let us introduce a scaling. We define 
$$H(\kappa)=\hp\otimes 1 +\kappa^21\otimes \hf+\kappa \hi,$$ 
where $\kappa>0$ is a scaling parameter. 
We shall outline a weak coupling limit in a heuristic level. 
Let ${\cal C}:=C([0,\infty);\RR^{dN})$.
It can be seen that 
\eq{br2}
(f\otimes \Omega, e^{-TH(\kappa)} g\otimes \Omega)=
\int_{{\cal C}\times\RR^{dN}} 
\ov{f(X_0)} g(X_t) e^{-\int_0^T V(X_s) ds} e^{W_\kappa}dP^x dx,
\en 
where 
$X_\cdot=(X_{1,\cdot},...,X_{N,\cdot}) \in {\cal C}
$, 
$dP^x$, $x\in\RR^{dN}$,  denotes the Wiener measure on ${\cal C}$ with 
$P^x(X_0=x)=1$, 
$$V(X_s):=\sum_{j=1}^N V_j(X_{j,s})$$ 
and 
\eq{br8}
W_\kappa:=\frac{1}{4}\sum_{i,j=1}^N \alpha_i\alpha_j 
\int_0^T ds\int_0^T dt \int_\BR {{\la_i(-k)}\la_j(k)}
\kappa^2 e^{-\kappa^2 |s-t|\omega(k)}
e^{-ik\cdot (X_{i,s}-X_{j,t})} dk.
\en 
Informally taking $\kappa\rightarrow \infty$ in \kak{br8},  
we see that the diagonal 
part of  $\int_0^T ds \int_0^T dt$  survives and the off diagonal part  
is dumped by factor 
$$\kappa^2 e^{-\kappa^2 |s-t|\omega(k)}
= \frac{
\omega(k)\kappa^2 e^{-\kappa^2|s-t|\omega(k)}}{\omega(k)}\sim\delta(s-t)\frac{1}{\omega(k)}.$$
Thus we have  
\eq{br3}
 W_\kappa\sim 
\frac{1}{4}\sum_{i,j=1}^N \alpha_i\alpha_j 
\int_0^T ds  \int_\BR \frac{{\la_i(-k)}\la_j(k)}
{\omega(k)}e^{-ik\cdot (X_{i,s}-X_{j,s})} dk
\en 
for a sufficiently large $\kappa$. 
Combining the right-hand side of \kak{br3} 
with $\int_0^T V(X_s) ds$ in \kak{br2}, 
we can derive 
the Feynman-Kac type path integral: 
\eq{br9}
\lim_{\kappa\rightarrow \infty}\kak{br2}
=
\int_{{\cal C}\times\RR^{dN}} 
\ov{f(X_0)} g(X_t) e^{-\int_0^T [
V(X_s) +\eff(X_s)+G]ds}dP^x dx,
\en 
where 
\eq{vv}
\eff(x):=\eff(x_1,...,x_n):=-\frac{1}{4}\sum_{i\not=j}^N \alpha_i\alpha_j
\int_\BR \frac{{\la_i(-k)}\la_j(k)}{\omega(k)} e^{-ik(x_i-x_j)}dk
\en 
and 
$$G:=-\frac{1}{4}\sum_{j=1}^N \int_\BR \frac{\hat\lambda_j(-k)\hat\lambda_j(k)}{\omega(k)}dk.$$
Note that when ${\rm supp}\la_i\cap{\rm supp}\la_j=\emptyset$, $i\not =j$, 
the effective potential $\eff$  vanishes. 
Heuristic arguments mentioned above can be operator theoretically 
established. 
Let 
$$\heff:=\sum_{j=1}^N\left(-\frac{1}{2m_j} \Delta_j+V_j \right) +\eff.$$
\begin{proposition}
\label{br6}
It follows that 
$$\slim_{\kappa\rightarrow\infty} 
e^{-t H(\kappa)}  
=e^{-t(\heff+G)} \otimes P_\Omega,$$
where 
$P_\Omega$ denotes the projection onto the Fock vacuum. 
\end{proposition}
See e.g., \cite{h5,h6} for details.  
Intuitively Proposition \ref{br6} suggests that 
$H(\kappa)\sim \heff+G$ for a sufficiently large $\kappa$. 
Then if $\heff$ has a ground state, 
$H(\kappa)$ also may have a ground state. 
This is actually proved by checking binding conditions 
introduced by \cite{gll} 
under the assumption that $\heff$ has a ground state.
This is an idea in this paper.  
\begin{remark}
Probabilistically through a  weak coupling limit, 
one can derive a Markov process from 
a non Markov process.  
The family of measures $\mu_\kappa$, $\kappa>0$,  
on ${\cal C}$ 
is given by 
\eq{br1}
\mu_\kappa(dX)=e^{-\int_0^t V(X_s) ds} 
e^{W_\kappa}  dP^x.
\en 
The double integral $W_\kappa$ in  \kak{br1}
breaks a Markov property of $(X_s)_{s>0}$ and 
$$T_{\kappa,s}:f\longmapsto 
\int _{{\cal C}} f(X_s)\mu_\kappa(dX),\quad\kappa<\infty,$$
does not define a semigroup on $L^2(\RR^{dN})$.  
The Markov property revives, however,  
as $\kappa\rightarrow \infty$,  and we have $T_{\infty, s}=e^{-s(\heff+G)}$.
\end{remark}

\subsection{Effective Hamiltonians and enhanced bindings}
Typical example of $\eff$ is a three dimensional  
$N$-body smeared Coulomb potential:  
$$\eff(x_1,...,x_N)=-\frac{1}{8\pi}\sum_{i\not=j}^N
 \frac{\alpha_i\alpha_j}{|x_i-x_j|}  \varpi (|x_i-x_j|),$$
where $\varpi (|x|)>0$ holds for a sufficiently small $|x|$. See \kak{F}.  
For this case it is determined by signs of 
$\alpha_1,...,\alpha_N$ 
whether  $\eff$ is attractive or repulsive for sufficiently
 small $|x_i-x_j|$.    
We can see from \kak{vv} that  an identical sign of coupling constants and
 ${\rm supp}\la_i\cap{\rm supp}\la_j\not=\emptyset$, $i\not =j$, 
derive attractive effective potentials and  enhances  binding of the system. 
Notice that although in the case of $N=1$ 
the enhanced binding in the Pauli-Fierz Hamiltonian occurs 
\cite{cvv, hvv, hisp1}, 
the effective potential \kak{vv}
disappears and then no enhanced binding in the Nelson model. 
This is a remarkable discrepancy 
between a nonrelativistic quantum electrodynamics and the Nelson model.

It is shown in e.g., \cite{dg, hi, lms} that 
the Nelson Hamiltonian with no infrared cutoff, $\la/\omega\not\in \LR$, 
has no ground state. 
So we do not discuss the infrared problem
and  assume that 
$\la/\omega\in\LR$. 
Moreover since we take the Boltzmann statistics for $N$ particles, it is established 
in \cite{bfs2} that the ground state is unique if it exists.  
Then we concentrate our discussion to showing  the existence of a ground state of $H$. 
Systems including the Fermi statistics will be discussed somewhere.  
We unitarily transform $H(\kappa)$ to a self-adjoint operator of the form 
\eq{U}
\heff\otimes 1+ \kappa^2 1\otimes \hf +H'(\kappa).
\en 
See Proposition \ref{1}.  
It is checked that under some condition $\heff$ has a ground state for 
$\alpha_j$'s  
with $0<\alpha_c<|\alpha_j|$, $j=1,\ldots,N$, for some $\alpha_c$, 
which suggests that for a sufficiently large $\kappa$,  $H(\kappa)$ also has a ground state for 
$\alpha_j$ with 
$\alpha_c<|\alpha_j|<\alpha_c(\kappa)$, $j=1,...,N$,  for some $\alpha_c(\kappa)$.  
Note that 
we do not assume the existence of ground states of $\hp$, namely $H(\kappa)$ with 
$\alpha_1=\cdots=\alpha_N=0$ may have no ground state.   
We show the existence of a ground state 
by checking  the binding condition \cite{gll}   in Proposition \ref{L2} 
for \kak{U}.

 If  there is no interaction between  particles, 
the $j$-th particle is influenced only 
 by the potential $V_j$. 
In this case, a shallow external potential 
$\sum_{j=1}^N V_j$ can not trap these
 particles. 
But if these particles attractively interact through an effective potential 
derived from a scalar bose field, 
particles close up and behave just like as one particle  
with mass $\sum_{j=1}^N m_j$. 
This \textit{one particle} may feel the force 
$-\sum_{j=1}^N \nabla_{x_j} V_j$.
If $N$ is large enough, this \textit{one particle} 
feels  $\sum_{j=1}^N  V_j$ strongly, and finally 
it will be trapped. 
In Section 3, we will justify this intuition.

This paper is organized as follows. 
In Section 2 the Nelson model and its 
scaled one is introduced and show the main results. 
The proof of the main theorem is also given. 
Section 3 is devoted to 
giving  examples of $\eff$ and $V_j$'s. 
Finally in Appendix A 
we show some fundamental facts on approximation 
of the bottom of the essential spectrum 
of Schr\"odinger operators.

\section{The main results and its proof}
\subsection{Statements and results}
Throughout this paper we assume (L) below:

\begin{itemize}
\item[ \bf{(L)} ]  For all $j=1,...,N$, (i),(ii),(iii) and (iv) are fulfilled. 
\item[ (i) ] 
$\la_j(-k)=\ov{\la_j(k)}$ and $\la_j\in\LR$, $\la_j/\sqrt\omega\in\LR$.
\item[(ii)]
 There exists an open set $S\subset \BR$ such that 
$\bar{S}=\supp \la_j $ and 
$\la_j\in C^1(S)$. 
\item[(iii)]
 For all $R>0$, $S_R:=\{k\in S| |k|<R\}$ has a cone property.
\item[(iv)]
 For all $p\in [1,2)$ and all $R>0$, $|\nabla_k \lambda_j|\in L^p(S_R)$.
\end{itemize}
\begin{remark}
(i) in (L) guarantees that $\hi$ is a symmetric operator. 
In the proof of Proposition \ref{L2} below, (ii)-(iv) in (L) are used. 
In order to show the existence of a ground state, 
we applied a method invented in  \cite{gll}.
Precisely, we used the photon derivative bound and 
the Rellich-Kondrachov theorem.
The conditions (ii)-(iv) are required to verify these procedures. 
See \cite{sasa} for details. 
In \cite{sasa} the dimension of the particle space equals
 three, but one can justify Proposition \ref{L2} in the $dN$-dimensional 
case. 
\end{remark}

Let $D(K)$ denote the domain of $K$. 
It is well known and easily proved that  $H$  is self-adjoint on 
$D(H):=D(\hp\otimes 1)\cap D(1\otimes \hf)$ and bounded from below 
for an arbitrary 
$\alpha_j\in\RR$, $j=1,...,N$, by the Kato-Rellich theorem with 
the inequality 
$$\|\hi\Psi\|\leq \epsilon \|H_0\Psi\|+b_\epsilon 
\|\Psi\|,\ \ \ \Psi\in D(H_0),$$
for an arbitrary $\epsilon>0$.  
It is also true that $H(\kappa)$ is self-adjoint on $D(H)$ for all $\kappa>0$. 

Assumptions (V1) and (V2) are introduced:

\begin{itemize}
\item[ \bf{(V1)} ] {\it There exists $\alpha_c>0$ such that 
$\inf\sigma(\heff)\in \s_{\rm disc}(\heff)$ 
for $\alpha_j$  with $|\alpha_j|>\alpha_c$, $j=1,...,N$.}
\item[ \bf{(V2)} ] {\it  $V_j(-\Delta+1)^{-1}$, $j=1,\ldots,N$,  
are compact. 
}
\end{itemize}
The main theorem is stated below. 
\bt{main}
Let  $\la_j/\omega\in\LR$, $j=1,...,N$,  and assume (L),(V1) and (V2). 
Fix a sufficiently large $\kappa>0$. 
Then for $\alpha_j$  
with $\alpha_c<|\alpha_j|<\alpha_c(\kappa)$, $j=1,...,N$, 
$H(\kappa)$ has a ground state, where $\alpha_c(\kappa)$ is a constant but 
possibly infinity. 
\et
The scaling parameter $\kappa$ in Theorem \ref{main} can be regarded  
as a dummy and 
absorbed into  $m_j$'s, $V_j$'s and $\la_j$'s. 
Let $\kappa$ be sufficiently large. 
Define 
$$\hat H:=\sum_{j=1}^N\left(-\frac{1}{2\hat m_j}\Delta_j+ \hat V_j\right)\otimes 1+
\sum_{j=1}^N \alpha_j \hat \phi_j+1\otimes \hf,$$
where 
$\hat m_j=m_j\kappa^2$, $\hat V_j=V_j/\kappa^2$ 
and $\hat \phi_j$ is defined by $\phi_j$ with $\la_j$ replaced by $\la_j/\kappa$. 
\bc{main2}Let  $\la_j/\omega\in\LR$, $j=1,...,N$,  and assume (L),(V1) and (V2). 
 Then 
$\hat H$ has a ground state for 
$\alpha_c<|\alpha_j|<\alpha_c(\kappa)$, $j=1,...,N$.
\ec
\proof 
We have $\kappa^{-2} H(\kappa)=\hat H$. 
Then by Theorem \ref{main}, $\hat H$ has a ground state. 
\qed

\subsection{Proof of Theorem \ref{main}}
Let $\la_j/\omega\in\LR$, $j=1,...,N$,  and define the unitary operator $T$ on $\hhh$ by 
$$T:=\exp\lk -i\sum_{j=1}^N\frac{\alpha_j}{\kappa}\pi_j\rk,$$
where 
$\d \pi_j:=\int_{\RR^{dN}}^\oplus \pi_j(x_j) dx$
with 
$$\pi_j(x):=\frac{i}{\sqrt2} 
\int_\BR \lk 
\add(k) e^{-ikx}\frac{\la_j(-k)}{\omega(k)}-a(k)e^{ikx}\frac{\la_j(k)}{\omega(k)} \rk dk.$$
\bp{1} $T$ maps $D(H)$ onto itself and 
\begin{eqnarray*}
&&\hspace{-0.5cm}
 T^{-1} H(\kappa) T\\
&&\hspace{-0.5cm} =\sum_{j=1}^N \lkk 
\frac{1}{2m_j}\lk -i\nabla_j\otimes 1 -\frac{\alpha_j}{\kappa}\tilde\phi_j\rk^2+V_j\otimes 1 
-\frac{\alpha_j^2}{2}
\|\la_j/\sqrt\omega\|^2\rkk+ \kappa ^2 1\otimes  \hf+\eff\otimes 1\\
&&\hspace{-0.5cm}=\heff\otimes 1+\kappa^2 1\otimes\hf+H'(\kappa),
\end{eqnarray*}
where 
$\d \tilde\phi_j:=\int_{\RR^{dN}}^\oplus \tilde\phi_j(x_j) dx$ with 
$$\tilde\phi_j(x):= \frac{1}{\sqrt 2}\int_\BR  k \lk \add(k) e^{-ikx}
\frac{\la_j(-k)}{\omega(k)}+
a(k) e^{ikx} \frac{\la_j(k)}{\omega(k)}\rk dk $$
and 
$$H'(\kappa)=\sum_{j=1}^N\lkk 
\frac{1}{\kappa}\frac{\alpha_j}{2m_j}((-i\nabla_j\otimes 1)\tilde\phi_j+\phi_j(-i\nabla_j\otimes 1))+
\frac{1}{\kappa^2} \frac{\alpha_j^2}{2m_j}\tilde\phi_j^2- 
\frac{\alpha^2_j}{2}\|\la_j/\sqrt\omega\|^2\rkk.$$
\ep 
\proof 
It is a fundamental identity. We omit the proof. 
\qed
Let us set 
$\N:=\{1,...,N\}$. 
For $\beta\subset \N$, we define
\begin{eqnarray*}
&& H^0(\beta)=H^0(\beta,\kappa) := \sum_{j\in \beta}
           \frac{1}{2m_j} \left(-i\nabla_j\otimes 1 -\frac{\alpha_j}{\kappa}\tilde\phi_j\right)^2
           + \kappa^21 \otimes H_f
	    + V_\mathrm{eff}(\beta)\otimes 1, \\
&& V_\mathrm{eff}(\beta) :=\lkk 
\begin{array}{ll} 
\displaystyle -\frac{1}{4}\sum_{i,j\in \beta, i\neq j}\alpha_i\alpha_j
 \int_{\BR}\frac{{\hat{\lambda}_i(-k)}\hat{\lambda}_j(k)}{\omega(k)}
 e^{-ik\cdot (x_i-x_j)}dk,&  |\beta|\geq 2,\\
0,&  |\beta | =0,1,\end{array}\right.
\\
&& 
H^V(\beta)=H^V(\beta,\kappa) := H^0(\beta) + \sum_{j\in \beta} V_j\otimes 1.
\end{eqnarray*}
 Simply we set 
$ H^V:=H^V(\N )$. 
 $H^V=H(\kappa)-\sum_{j=1}^N \alpha_j^2\|\la_j\|^2/4$ has ground states 
if and only if 
$H(\kappa)$ does, since 
$\sum_{j=1}^N \alpha_j^2\|\la_j\|^2/4$ is a fixed number.
In what follows our investigation is focused on 
showing  the existence of ground state of $H^V$.   
The operators $H^0(\beta)$ and $H^V(\beta)$ 
are self-adjoint operators acting on 
$L^2({\RR}^{d|\beta|})\otimes\mathcal{F}$.
We set 
\begin{align*}
\begin{array}{ll}  E^V(\kappa):=\is(H^V),& 
 E^V(\kappa,\beta) := \inf \sigma(H^V(\beta)),\\
 E^0(\kappa,\beta) := \inf \sigma(H^0(\beta)), & 
E^V(\kappa,\emptyset) := 0.\end{array}
\end{align*}
The lowest two cluster threshold $\Sigma^V(\kappa)$ is defined by
\begin{align*}
 \Sigma^V(\kappa):= \min\{E^V(\kappa,\beta) + E^0(\kappa,\beta^{\mathrm{c}})| 
                            \beta\subsetneqq \N \}.
\end{align*}
To establish the existence of ground state of $H(\kappa)$, we use 
the next proposition:
\begin{proposition}[\cite{gll}]{\label{L2}}
Let 
$  \Sigma^V(\kappa) - E^V(\kappa) >0.
$
Then  
$H(\kappa)$ has a ground state.
\end{proposition}
For $\beta\subset\N $, we set 
\begin{eqnarray*}
\begin{array}{ll}
\d h^0(\beta) := -\sum_{j\in \beta} \frac{1}{2m_j}\Delta_j 
          +V_\mathrm{eff}(\beta),  &
h^V(\beta) := h^0(\beta) + \sum_{j\in \beta}V_j, \\
\cE^0(\beta) := \inf\sigma(h^0(\beta)),& 
\cE^V(\beta) := \inf\sigma(h^V(\beta)), 
\end{array}
\end{eqnarray*}
where 
$ h^0(\emptyset) :=0$ and $ h^V(\emptyset) := 0$. 
Furthermore 
we simply put 
\eq{z1}
h^V:=h^V(\N)=\heff,\quad 
\cE ^V:=\inf\sigma(h^V).
\en 
We define the lowest two cluster threshold for $h^V$ by 
\eq{th}
\Xi^V := \min\{\cE^V(\beta)+ \cE^0(\beta^\mathrm{c})|
\beta\subsetneqq \N \}
\en
and we set 
$$\eff_{ij}(x):=
-\frac{1}{4}\alpha_i\alpha_j\int_{\RR^d} \frac{\la_i(-k)\la_j(k)}{\omega(k)}
e^{-ik\cdot x}
dk,\ \ \ i\not =j.$$
\bl{v}
Potentials $\eff_{ij}$, $i,j=1,...,N$, are relatively compact with respect to the 
$d$-dimensional Laplacian.
\el
\proof 
Since $\la_i\la_j/\omega\in L^1(\BR)$, $i,j=1,...,N$, we can see that 
$\eff_{ij}(x)$ is continuous in $x$ and 
$\lim_{|x|\rightarrow \infty}\eff_{ij}(x)=0$ by the Riemann-Lebesgue 
theorem. 
In particular $\eff_{ij}$ is relatively compact with respect to the $d$-dimensional Laplacian. 
\qed

We want to estimate $\is_{\rm ess}(\heff)$. 
For Hamiltonians with the center of mass motion removed, the bottom of the essential spectrum is 
estimated by HVZ theorem. By extending the IMS localization argument to a quantum field model, 
in \cite{gll} the lowest two cluster threshold 
of a Hamiltonian interacting with a quantized field (the Pauli-Fierz model) is shown.  
The following lemma is a simplified version of \cite{gll}, since no interaction with a quantized radiation field exists. 
For a self consistency of this paper we give an outline of a  proof. 
\begin{lemma}\label{hvz}
Assume (V2). 
Then 
$ \sigma_\mathrm{ess}(H_\mathrm{eff})~=~[\Xi^V,\infty) $. 
\end{lemma}
\begin{proof}
We may assume that $V_i, \eff_{ij} \in C_0^\infty(\BR)$ 
by Proposition \ref{A2}. 
Then  
there exists a normalized sequence  $\{g_n\}_n\subset 
 C_0^\infty(\RR^{dN})$ 
such that ${\rm supp} g_n\subset 
\{x\in \RR^{dN}| V_i(x)=0, \eff_{ij}(x_i-x_j)=0, i,j=1,...,N\}$ 
and 
$(g_n, h^V(\beta)g_n)=(g_n, \sum_{j\in\beta}(-\Delta /2m_j) g_n)\rightarrow 0$ 
as $n\rightarrow \infty$. 
Then we have 
\eq{B7}
{\cal E}^V(\beta)+{\cal E}^0(\beta^c)\leq 0.
\en
Let $\tilde{j}_\beta\in C^\infty(\mathbb{R}^d)$, $\beta\in \N$,  
be 
a Ruelle-Simon partition of unity 
\cite[Definition 3.4]{cfks}, which satisfy 
(i)-(v)  below:\\
(i) $ 
	\sum_{\beta\subseteq \N } \tilde{j}_\beta(x)^2 =1,
       $\\
(ii)  $\d 		\tilde{j}_\beta(Cx) = \tilde{j}_\beta(x)$ 
for $|x|=1$, $C\geq 1$ and $\beta\neq \N ,
	       $ \\
(iii)     
$\d 
  \mathrm{supp}\> \tilde{j}_\beta
  \subset \{x\in\mathbb{R}^d| 
  \min_{i\in\beta,j\in\beta^\mathrm{c}}\{|x_i-x_j|,|x_j|\}\geq c|x|\}
$ for some $c>0$,\\
(iv) $\d 
              \tilde{j}_\beta(x) = 0$ for $|x|<\frac{1}{2}$ 
and $\beta\neq \N $, \\
(v) $\tilde{j}_{\N }$ has a compact support.

For a constant $R>0$ we put 
$j_\beta(x) := \tilde{j}_\beta(x/R)$. 
Note that for each $\beta\subset \N $, 
$$H_\mathrm{eff}=h^V(\beta)\otimes 1+1\otimes h^0(\beta^c)+
\underbrace{
\sum_{i\in\beta^c}1\otimes V_i(x_i)+
\sum_{\substack{i\in\beta,j\in\beta^\mathrm{c} 
\\ i\in\beta^\mathrm{c},j\in\beta}} \eff_{ij}(x_i-x_j)}_{=I_\beta}.$$
By the IMS localization formula \cite[Theorem 3.2 and p. 34]{cfks},  
we have 
$$
 H_\mathrm{eff} 
=
 j_{\N } H_\mathrm{eff} j_{\N } +
      \sum_{\beta\subsetneqq\N } j_\beta \left[
      h^V(\beta)\otimes 1+1\otimes h^0(\beta^\mathrm{c})      
       \right] j_\beta 
+\sum_{\beta\subsetneqq\N } j_\beta^2 I_\beta  
  - \frac{1}{2}\sum_{\beta\subseteqq\N }|\nabla j_\beta|^2.
   \label{small}
$$
Here 
we identify as
$L^2(\RR^{dN})\cong 
L^2(\RR^{d|\beta|})\otimes L^2(\RR^{d|\beta^c|})$. 
Since 
$j_{\N }^2  (\sum_{j=1}^N V_j+\eff)  $ and 
$\sum_{\beta\subsetneqq\N } j_\beta^2 I_\beta$ 
are relatively compact 
with respect to the $dN$-dimensional Laplacian 
by the property (iii) and (v), it is seen that  
\begin{eqnarray*}
&&
\hspace{-1cm}
\s_{\rm ess}(H_{\rm eff})\\
&&\hspace{-1cm}
=
\s_{\rm ess}\lk j_\N(-\half\sum_{j=1}^N\Delta_j) j_\N+
\sum_{\beta\subsetneqq\N } j_\beta \left[
      h^V(\beta)\otimes 1+1\otimes h^0(\beta^\mathrm{c})      
       \right] j_\beta 
     - \frac{1}{2}\sum_{\beta\subseteqq\N }|\nabla j_\beta|^2 \rk.
\end{eqnarray*}
We have 
$$
\sum_{\beta\subsetneqq\N } j_\beta \left[
      h^V(\beta)\otimes 1+1\otimes h^0(\beta^\mathrm{c})      
       \right] j_\beta 
 \geq  \sum_{\beta\subsetneqq\N } 
   (\cE^V(\beta)+\cE^0(\beta^\mathrm{c}))j_\beta^2.
$$
By (ii) and (v),  
$$
\Big\| 
\frac{1}{2} \sum_{\beta\subseteqq\N }|\nabla j_\beta|^2\Big\|
 \leq \frac{C}{R^2}
$$ with some constant $C$ independent of $R$. 
Hence we obtain
that $$
 \inf\sigma_\mathrm{ess}(H_\mathrm{eff})
 \geq 
 \min_{x\in\mathbb{R}^d} \sum_{\beta\subsetneqq \N }
 (\cE^V(\beta)+ \cE^0(\beta^\mathrm{c}))
 j_\beta(x)^2 - \frac{C}{R^2}
 \geq 
 \Xi^V -\frac{C}{R^2}
$$
for all $R>0$. Here we used (i) and \kak{B7}.
Thus 
$
\sigma_\mathrm{ess}(H_\mathrm{eff}) \subset  [\Xi^V,\infty)$ follows. 
Next we shall prove the reverse inclusion 
$\sigma_\mathrm{ess}(H_\mathrm{eff}) \supset  [\Xi^V,\infty)$.
Fix $\beta\subsetneqq  \N $. 
Let $\{\psi_n^V \}_{n=1}^\infty\subset C_0^\infty(\mathbb{R}^{d|\beta|})$
 be a minimizing sequence of $h^V(\beta)$ so that 
\begin{align*}
 \lim_{n\to\infty}
 \|(h^V(\beta)-\cE^V(\beta))\psi_n^V\|
 =0, \quad \|\psi_n^V\|=1.
\end{align*}
and $\{\psi_n^0\}_{n=1}^\infty\subset 
C_0^\infty(\mathbb{R}^{d|\beta^\mathrm{c}|})$  a normalized 
sequence such that
\begin{align}
 \lim_{n\to\infty}
 \| (h^0(\beta^\mathrm{c})-\cE^0(\beta^\mathrm{c})-K)\psi_n^0\|=0,
 \label{minim2}
\end{align}
where $K\geq 0$ is a constant.
Note that since 
$\s(h^0(\beta^c))=[{\cal E}^0(\beta^c),\infty)$, 
$\psi_n^0$ such as \kak{minim2} exists.
By the translation invariance of $h^0(\beta^\mathrm{c})$, 
for any function $\tau_\cdot :{\Bbb N}\to \mathbb{R}^d$ the translated sequence 
$\psi_n^0(x_{j_1}-\tau_n,\ldots ,x_{j_{|\beta^\mathrm{c}|}}-\tau_n)$ also satisfies 
\kak{minim2}.
Let $R_n>0$ be a constant satisfying 
\begin{align*}
 \mathrm{supp}\> \psi_n^V \subset 
 \{x=(x_{j_1},\cdots,x_{j_{|\beta|}})\in \mathbb{R}^{d|\beta|}| |x_{j_i}|<R_n,
 \, j_i\in\beta, i=1,...,|\beta|\}.
\end{align*}
We take $\tau$  such that
\begin{align*}
& \mathrm{supp}\>\psi_n^0(\cdot-\tau_n,\cdots,\cdot-\tau_n) \\
&\subset
\{x=(x_{k_1},\cdots,x_{k_{|\beta^c|}})\in \mathbb{R}^{d|\beta^c|}| 
|x_{k_i}|\geq R_n+n,
 \, k_i\in\beta^c,i=1,...,|\beta^c|\}.
\end{align*}
We set
$
 \Psi_n(x_1\cdots x_N) = \psi_n^V(x_{j_1}\cdots x_{j_{|\beta|}})
\otimes \psi_n^0(x_{k_1}-\tau_n\cdots x_{k_{|\beta^c|}}-\tau_n)
\in L^2(\RR^{dN})$.
Then, for all $i,j$ with $i\in\beta, \, j\in\beta^\mathrm{c}$, we have
\begin{align*}
 \| \eff_{ij}(x_i-x_j)\Psi_n\| &\leq \sup_{x\in\mathbb{R}^d,|x|>n}
|\eff_{ij}(x)|
 \to 0, \quad (n\to\infty), \\
 \|V_j(x_j)\Psi_n\| &\leq \sup_{x\in\mathbb{R}^d, |x|\geq R_n+n}|V_j(x)|
 \to 0, \quad (n\to\infty).
\end{align*}
Hence, by a triangle inequality, we have that
\begin{align*}
 \|(H_{\mathrm{eff}}-\cE^V(\beta)-\cE^0(\beta^{\mathrm{c}})-K)\Psi_n\| \to 0, \quad 
(n\to\infty).
\end{align*}
Therefore $[{\cal E}^V(\beta)+{\cal E}^0(\beta^c)+K,\infty)
\subset \s(H_{\rm eff})$.
Since $\beta\subsetneqq \N$ and $K>0$ are arbitrary, 
 $ [\Xi^V,\infty) \subset \sigma_\mathrm{ess}(H_\mathrm{eff})$ follows. 
Thus the proof is complete.
\qed
\end{proof}
We define 
\begin{equation*}
  \Delta_\mathrm{p}(\alpha_1,...,\alpha_N ) :=\Xi^V - \cE^V.
\end{equation*}

\begin{corollary}{\label{L3}}
 Assume (V1) and (V.2). 
 Then $\Delta_\mathrm{p}(\alpha_1,...,\alpha_N )>0$ follows 
for $\alpha_j$  with 
$|\alpha_j|>\alpha_c$, $j=1,...,N$. 
\end{corollary}
\proof 
Since $\is_{\rm ess}(\heff)=\Xi^V$  by Lemma \ref{hvz} and 
$\is(\heff)\in \s_{\rm disc}(\heff)$ by (V1), 
the corollary follows from $\Delta_P(\alpha_1,...,\alpha_N )
=\is_{\rm ess}(\heff)-\is(\heff)>0$. 
\qed

\begin{lemma}{\label{L4}} For an arbitrary $\kappa>0$,  it follows that 
$\Sigma^V(\kappa)\geq \Xi^V$. 
\end{lemma}
\proof 
It is well known that $H^V(\beta)$ can be realized as a self-adjoint operator 
on a Hilbert space $\hhh_Q=L^2(\RR^{|\beta|d})\otimes L^2(Q,d\mu)$ with some 
measure space $(Q, \mu)$, which is called 
a Schr\"odinger representation. 
It is established that 
$$(\Psi, e^{-tH^V(\beta)}\Phi)_{\hhh_Q}\leq (|\Psi|, 
e^{-t(h^V(\beta)\otimes1+\kappa^2 1\otimes  \hf)}|\Phi|)_{\hhh_Q}.$$ 
Hence for any $\beta\subset\N $, it follows that 
$\is(h^V(\beta)\otimes 1+\kappa^2 1\otimes \hf )\leq \is(H^V(\beta))$. 
Since $\is(\hf)=0$ and $\is(h^V(\beta)\otimes 1+\kappa^2 1\otimes \hf )=\is(h^V(\beta))$,  the lemma 
follows from  the definition of lowest two cluster thresholds. 
\qed

\begin{lemma}{\label{L5}}
 Assume (V1). Then
$  E(\kappa) \leq \cE^V
  + \kappa^{-2}\sum_{j=1}^N \alpha_j^2\|\la_j\|^2/(4m_j)$ 
for $\alpha_j$ with 
$|\alpha_j|>\alpha_c$, $j=1,...,N$.
\end{lemma}
\begin{proof}
By (V1), $H_\mathrm{eff}$ has a normalized ground state $u$ for $\alpha_j$ with 
$|\alpha_j|>\alpha_c$, $j=1,...,N$. 
Set $\Psi:= u\otimes \Omega$.
Then 
\begin{eqnarray*}
 E(\kappa) 
&\leq & (u, H_\mathrm{eff}u )
     + \sum_{j=1}^N \frac{\alpha_j}{2m_j\kappa}
       2\Re ( i\nabla_j\Psi, \tilde\phi_j\Psi)
     + \sum_{j=1}^N \frac{\alpha_j^2}{2m_j\kappa^2}
       \| \tilde\phi_j\Psi\|^2 \\
   &=&
      \cE^V +
      \sum_{j=1}^N \frac{\alpha_j^2}{4m_j\kappa^2} \|\lambda_j\|^2.
\end{eqnarray*}
Then the lemma follows. 
\qed
\end{proof}

\textit{Proof of Theorem \ref{main}}\\
By Lemmas \ref{L4} and \ref{L5}, we have
$$
 \Sigma^V(\kappa) -E(\kappa)
 \geq \Xi^V -\cE^V- \sum_{j=1}^N \frac{\alpha_j^2}{4m_j\kappa^2} \|\lambda_j\|^2 \\
 = \Delta_\mathrm{p}(\alpha_1,...,\alpha_N ) 
      - \sum_{j=1}^N \frac{\alpha_j^2}{4m_j\kappa^2} \|\lambda_j\|^2. 
$$
Note that $\Delta_p(\alpha_1,...,\alpha_N )>0$ is continuous in $\alpha_1,...,\alpha_N$. 
Then for a sufficiently large $\kappa$, 
we can obtain that 
there exists $\alpha_c(\kappa)>\alpha_c$ such that 
for $\alpha_c<|\alpha_j|<\alpha_c(\kappa)$, $j=1,...,N$, 
$\Sigma^V(\kappa)-E(\kappa)>0$. 
Thus  $H(\kappa)$ has 
a ground state for such $\alpha_j$'s by Proposition~\ref{L2}. 
\qed

\section{Examples}
\subsection{Example of effective potentials}
\label{ex}
The typical example of ultraviolet cutoff function is of the form $\la_j=\rh_j/\sqrt\omega$, 
$j\in \N $, 
with rotation invariant nonnegative functions $\rh_j$. 
In this case $\eff(x_1,...,x_N)=\sum_{i\not=j}\alpha_i\alpha_j
\eff_{ij}(x_i-x_j)$ 
satisfies that 
(1) $\eff_{ij}$ is continuous, 
(2) $\lim_{|x|\rightarrow\infty}V_{ij}(x)=0$ and 
(3) $\eff_{ij}(0)<\eff_{ij}(x)$ for all $x\in \BR$ but $x\not=0$.
More explicitly 
effective potential $\eff$ is given by 
\begin{eqnarray*}
&&
\eff(x_1,\cdots,x_N)=-\frac{1}{4}\sum_{i\not= j}^N\alpha_i\alpha_j 
\int_{\BR}\frac{\rh_i(-k)\rh_j(k)}{\omega(k)^2} e^{-ik\cdot (x_i-x_j)} dk\\
&&=
-\frac{1}{4}\sum_{i\not= j}^N \alpha_i\alpha_j
\frac{\sqrt{(2\pi)^d}  }{|x_i-x_j|^{(d-1)/2}}\int_0^\infty\frac{r^{(d-1)/2}}{r^2}\rh_i(r)\rh_j(r)\sqrt{r|x_i-x_j|}J_{(d-2)/2}(r|x|) 
dr.
\end{eqnarray*}
Here 
$J_\nu$ is the Bessel function: 
$$J_\nu(x)=(x/2)^\nu\sum_{n=0}^\infty\frac{(-1)^n}{n!\Gamma(n+\nu+1)}(x/2)^{2n}$$ where $\Gamma$ 
denotes the Gamma function. 
In the case of $d=3$, and 
$$\rh_j(k)=\lkk\begin{array} {ll}
0&|k|<\kappa,\\ 
1/\sqrt{(2\pi)^3}& \kappa<|k|<\Lambda,\\ 
0&|k|\geq \Lambda,\end{array}\right.$$
we see that 
\eq{F}
\eff(x_1,\cdots,x_N)= -\frac{1}{8\pi^2} \sum_{i\not= j}^N 
\frac{\alpha_i\alpha_j}{|x_i-x_j|}\int_{\kappa|x_i-x_j|}^{\Lambda|x_i-x_j|} 
\frac{\sin r}{r}dr.
\en 
\subsection{Example of $V_j$'s}
We give an example of $V_1,\cdots,V_N$ satisfying assumption (V1). 
Assume simply that $V_1=\cdots=V_N=V$, 
 $\alpha_1=\cdots=\alpha_N=\alpha$, $\la_1=\cdots=\la_N=\la$ and 
$m_1=\cdots=m_N=m$. 
Then 
$\eff_{ij}=W$ for all $i\not= j$. 
Let 
\begin{equation*}
 h^{V}(\alpha) := 
 \sum_{j=1}^N \left(-\frac{1}{2m}\Delta_j +V(x_j) \right)
 + \alpha^2 \sum_{j\neq l}^N \WWW (x_j-x_l),
\end{equation*}
which acts on $L^2(\mathbb{R}^{dN})$.
We assume (W1)-(W3) below:\\
\begin{description}
\item[(W1)]
$V$ is  relatively compact with respect to 
the $d$-dimensional Laplacian $\Delta$, and 
$
 \sigma(-(\Delta /2m)+V) = [0,\infty)$.
\item[(W2)]
$\WWW $ satisfies that 
$\d 
 -\infty < \WWW (0)= \essinf _{|x|<\epsilon}\WWW (x) < 
 \essinf_{|x|>\epsilon} \WWW (x) $ for all $\epsilon>0$. 
\item[(W3)]
$\is( -(\Delta/(2Nm) 
 + N V )\in \s_{\rm disc} 
( -(\Delta/(2Nm) 
 + N V )$. 
\end{description}
\begin{remark}
Note that examples of $\eff$ given in subsection \ref{ex} satisfies (W2). 
The condition (W1) means that the external potential $V$ is shallow and 
the non-interacting Hamiltonian $h^{V}(0)$ has no negative energy bound state.
\end{remark}
When $W=0$, (W1) implies that 
each particle independently behaves and 
is not trapped. 
When $W\not=0$, $W$ closes up $N$ particles and 
they behave  as {\it one particle} with mass $Nm$.
The \textit{one particle} may feel the force $-N\nabla V$ and 
 be trapped  by $N V$. 
The following theorem justifies this heuristic argument. 

\begin{theorem}{\label{Tex}}
 Assume (W1)-(W3).
Then, there exists $\alpha_c>0$ such that for all $\alpha$ with $|\alpha|>\alpha_c$,  
$\is(h^{V}(\alpha))\in \s_{\rm disc}(h^V(\alpha))$.
\end{theorem}
To prove Theorem \ref{Tex} we need  several lemmas. 
For  $\beta\subset  \N$,  we define
\begin{align*}
  h^0(\alpha,\beta) &:= -\frac{1}{2m}\sum_{j\in\beta}\Delta_{j}
                   +\alpha^2 \sum_{\substack{j,l\in\beta\\ j\neq l}}\WWW (x_j-x_l),\quad 
  h^V(\alpha,\beta) := h^0(\alpha,\beta) +\sum_{j\in\beta}V(x_j), \\
  \cE^0(\alpha,\beta) &:= \inf\sigma(h^0(\alpha,\beta)), \quad  \cE^V(\alpha,\beta) := 
\inf\sigma(h^V(\alpha,\beta)), 
\end{align*}
where 
$\cE^V(\alpha,\emptyset):=0$ and 
$\cE^0(\alpha,\emptyset):=0$. 
Simply we set 
${\cal E}^V(\alpha,\N)={\cal E}^V(\alpha)$ 
and ${\cal E}^0(\alpha,\N)={\cal E}^0(\alpha)$.
Let $\Xi^V(\alpha) $ denote the lowest two cluster threshold of $h^V(\alpha)$ defined by \kak{th}. 
Then by  (W1) and Lemma \ref{hvz}, we have
\eq{sd}
 \sigma_\mathrm{ess}(h^V(\alpha))
 = [\Xi^V(\alpha),\infty).
\en
\begin{lemma}\label{ene}
 Let $\beta\subsetneqq \N $ but $\beta\neq\emptyset$. 
Then there exists  $\alpha'>0$ such that, 
for all $\alpha$ with $|\alpha|>\alpha'$,  
\eq{sas}
   \cE^0(\alpha) < \cE^V(\alpha,\beta)+ \cE^0(\alpha,\beta^c).
  \en
\end{lemma}
\begin{proof}
 Since $h^0(\alpha,\beta)/\alpha^2 $ and 
$h^V(\alpha,\beta)/\alpha^2$ converge to $\sum_{\substack{j,l\in\beta\\ j\neq l}}\WWW (x_j-x_l)$
 in the uniform resolvent sense, 
by (W2), one can show that
  \begin{equation*}
    \lim_{\alpha\to\infty}\frac{\cE^V(\alpha, \beta)}{\alpha^2}
   = \lim_{\alpha\to\infty}\frac{\cE^0(\alpha, \beta)}{\alpha^2}
   = |\beta|(|\beta|-1) \WWW (0).
  \end{equation*}
\end{proof}
Hence
\begin{align*}
  \lim_{\alpha\to\infty}\frac{\cE^0(\alpha)}{\alpha^2}
  &= N(N-1) \WWW (0), \\
  \lim_{\alpha\to\infty}\frac{\cE^V(\alpha,\beta)+\cE^0(\alpha,\beta^c)}{\alpha^2}
  &= \big\{(|\beta|(|\beta|-1) + |\beta^c|(|\beta^c|-1)\big\}
     \WWW (0) \\
  &= \big\{N(N-1) +2|\beta|(|\beta|-N) \big\} \WWW (0).
\end{align*}
Since 
$|\beta|(|\beta|-N)\leq -1$ and 
$\WWW (0)<0$ by (W2), 
we see that there exists  $\alpha'>0$ such that \kak{sas} holds for all $\alpha$ with 
$|\alpha|>\alpha'$. 
\qed

Let $X=(x_1,...,x_N)^t\in \RR^{dN}$ and $Y:=(x_c,y_1,\ldots,y_{N-1})^t$ be 
its  Jacobi coordinates:
\begin{align*}
  x_c := \frac{1}{N} \sum_{j=1}^N x_j, \quad  
 y_j := x_{j+1} - \frac{1}{j}\sum_{i=1}^j x_i, \quad j=1,...,N-1.
\end{align*}
Let $T\in {\rm GL}(N,\RR)$ be  such that 
$Y = TX$. 
{Note that
\begin{align*}
T&=\begin{bmatrix}
 \frac{1}{N}   &  \frac{1}{N}  & \frac{1}{N}   & \cdots&\cdots &\cdots & \frac{1}{N} \\
 -1            &  1            &  0            & \cdots & \cdots & \cdots & 0   \\
 -\frac{1}{2}  & -\frac{1}{2}  &  1            & 0 & 0  & \cdots & 0  \\
 -\frac{1}{3}  &  -\frac{1}{3} & -\frac{1}{3}  & 1      & 0 &\cdots & 0 \\
 \vdots        &    \vdots     & \vdots        & \cdots & \ddots& \cdots    & \cdots \\
 \vdots        &    \vdots     & \vdots        & \cdots & \cdots& \ddots    & \cdots \\
 -\frac{1}{N-1}&-\frac{1}{N-1} & -\frac{1}{N-1}& \cdots & \cdots& -\frac{1}{N-1} & 1
    \end{bmatrix}, \\
T^{-1}&=
\begin{bmatrix}
1 &-\frac{1}{2}&-\frac{1}{3}&-\frac{1}{4}&-\frac{1}{5}& \cdots&\cdots& -\frac{1}{N} \\
1 &\frac{1}{2}&-\frac{1}{3}&-\frac{1}{4}&-\frac{1}{5}&\cdots &\cdots& -\frac{1}{N} \\
1 &  0 &\frac{2}{3} & -\frac{1}{4} & -\frac{1}{5}&\cdots&\cdots& -\frac{1}{N}\\
1 & 0 & 0 & \frac{3}{4} & -\frac{1}{5} & -\frac{1}{6} & \cdots & -\frac{1}{N}\\
\vdots& \vdots&\cdots& \cdots&\ddots& \cdots&\cdots& \vdots \\
\vdots&\vdots&\cdots& \cdots&\cdots& \ddots&\cdots& \vdots \\
1 & 0 &\cdots& \cdots&\cdots&  0& \frac{N-2}{N-1}&  -\frac{1}{N} \\
1 & 0 & \cdots&\cdots&\cdots& \cdots&  0& \frac{N-1}{N}
\end{bmatrix}.
\end{align*}
$T$  induces 
the  unitary operator 
 $U:L^2(\mathbb{R}^{dN}_X)\to L^2(\mathbb{R}^{dN}_Y)$ defined 
by
$ (U\psi)(Y) := \psi(T^{-1}Y)$.
We have 
\begin{align*}
 & U h^0(\alpha) U^{-1} 
 = -\frac{1}{2Nm} \Delta_{x_c} 
   - \sum_{j=1}^N \frac{1}{2\mu_j}\Delta_{y_j}
      + \alpha^2\sum_{\substack{j\neq l}}^N \WWW (x_j(Y)-x_l(Y)), \\
 & U h^V(\alpha) U^{-1}
   = U h^0(\alpha) U^{-1}  + \sum_{j=1}^N V(x_j(Y)),
\end{align*}
where $\mu_j:= jm/(j+1)$ is a reduced mass and $x_j(Y):=(T^{-1}Y)_j$.
Let $k(\alpha)$ be $h^0(\alpha)$ with the center of mass motion removed:
\begin{align*}
  k(\alpha) := - \sum_{j=1}^N \frac{1}{2\mu_j}\Delta_{y_j}
      + \alpha^2\sum_{\substack{j\neq l}}^N \WWW (x_j(Y)-x_l(Y)).
\end{align*}
Set 
$\RR^{dN}=\RR^d_{x_c}\oplus\RR^{d(N-1)}_{y_1,...,y_{N-1}}:=
\chi_c\oplus\chi_c^\perp$. 
Since   $x_j(Y)-x_i(Y)$, $i,j=1,...,N-1$,  depend only 
on $y_1,\ldots, y_{N-1}\in \chi_c^\perp$, 
$k(\alpha)$ is a self-adjoint operator acting on 
$L^2(\chi_c^\perp)$. 
\begin{lemma}
  There exists  $\alpha''>0$ such that   $\is(k(\alpha))\in \s_{\rm disc}(k(\alpha))$ for all $\alpha$ with $|\alpha|>\alpha''$.
\end{lemma}
\begin{proof}
Assume  that $\lim_{|x|\to\infty}\WWW (x)=0$.
Let $\chi, \bar\chi \in C^\infty(\mathbb{R})$ be such that
$
   \chi(x)^2 + \bar\chi(x)^2 =1$ with 
$ \chi(x) =
   \begin{cases}
     1, \quad |x|<1, \\
     0, \quad |x|>2.
   \end{cases}$
For a parameter $R$, we set 
\begin{align*}
\chi_R(y_1)&:=\chi(|y_1|/R), \quad \bar\chi_R(y_1):=\bar\chi(|y_1|/R),
 \quad  y_1\in \mathbb{R}^d, \\
\theta_R(Y_1)&:= \chi(|Y_1|/2R), \quad \bar\theta_R(Y_1):= \bar\chi(|Y_1|/2R),
 \quad Y_1:=(y_2,\ldots,y_{N-1}) \in \mathbb{R}^{d(N-2)}.  
\end{align*}
By the IMS localization formula, we have 
\begin{eqnarray}
  k(\alpha) &=& \chi_R \theta_R k(\alpha) \theta_R \chi_R
             +\chi_R \bar\theta_R k(\alpha) \bar\theta_R \chi_R
             +\bar\chi_R k(\alpha) \bar\chi_R  \non \\ 
             & & \label{kk2} 
\underbrace{-\frac{1}{2}\chi_R^2 |\nabla\theta_R|^2   
               -\frac{1}{2}\chi_R^2 |\nabla\bar\theta_R|^2
              -\frac{1}{2}|\nabla\chi_R|^2
             -\frac{1}{2}|\nabla\bar\chi_R|^2}_{=B(R)}.
\end{eqnarray}
Here $B(R)$ 
is a bounded operator with 
$$\| B(R)\|
\leq \frac{C}{R^2},$$
where $C$ is  a constant independent of $R$. 
Since $\chi_R^2\theta_R^2 \alpha^2 \sum_{j\not= l}^N W(x_j(Y)-x_l(Y))$ is 
relatively compact with respect to $-\sum_{j=1}^N (2\mu_j)^{-1}\Delta_{y_j}$, 
we have 
$\s_{\rm ess}(k(\alpha))
=\s_{\rm ess}(k'(\alpha))$, 
where 
$$k'(\alpha)=
\chi_R \theta_R 
\Big(-\sum_{j=1}^N \frac{1}{2\mu_j}\Delta_{y_j}\Big)  \theta_R \chi_R
             +\chi_R \bar\theta_R k(\alpha) \bar\theta_R \chi_R
             +\bar\chi_R k(\alpha) \bar\chi_R  +B(R).$$
We have
\begin{align}
 k'(\alpha) \geq
 & \chi_R^2\bar\theta_R^2E
\lk k(\alpha)-\alpha^2 \WWW (x_2(Y)-x_3(Y))
 -\alpha^2 \WWW (x_3(Y)-x_2(Y))\rk  \label{K44}\\
 & + \chi_R^2\bar\theta_R^2\alpha^2 [
\WWW (x_2(Y)-x_3(Y))
   +  \WWW (x_3(Y)-x_2(Y))] \label{K45}\\
 & + \bar\chi_R^2 
E\lk k(\alpha) -\alpha^2 \WWW (x_1(Y)-x_2(Y))
     - \alpha^2 \WWW (x_2(Y)-x_1(Y))\rk  \label{K46}\\
 & +  \bar\chi_R^2 \alpha^2 [
\WWW (x_1(Y)-x_2(Y))
   +  \WWW (x_2(Y)-x_1(Y))]  \label{K47}\\
&-C/R^2\label{K43}.
\end{align}
Note that $y_1=x_2(Y)-x_1(Y)$ and $x_3(Y)-x_2(Y) = y_2-y_1/2$.
We have
\begin{align*}
  |(\ref{K45})| &\leq 2\sup_{\substack{y_1,y_2\\ |y_1|<2R, \, |y_2|>4R}}
                     \alpha^2 |\WWW (y_2-y_1/2)|
 \leq  2\alpha^2\sup_{|y|>3R} |\WWW (y)|, \\
  |(\ref{K47})| &\leq 2 \sup_{|y_1|>2R} \alpha^2 |\WWW (y_1)|.
\end{align*}
Since we assume that $\lim_{|x|\rightarrow \infty}\WWW (x)=0$, we obtain
that 
$ \lim_{R\to\infty} \|(\ref{K45})\| =0$ and 
$\lim_{R\to\infty} \|(\ref{K47})\| =0$. 
Thus, for all $R>0$ we have
\begin{eqnarray}
 \inf\sigma_{\mathrm{ess}}(k(\alpha)) &\geq &
 \inf_{Y\in\mathbb{R}^{d(N-1)}}[(\ref{K44})+(\ref{K46})] 
-\|\kak{K45}\|-\|\kak{K47}\|-C/R^2 \non \\
 &\geq & \min\{ E(k(\alpha)-\alpha^2 \WWW (x_1-x_2)
                    -\alpha^2 \WWW (x_2-x_1)), \non \\
 & &\label{y1}
\quad                 E(k(\alpha)-\alpha^2 \WWW (x_2-x_3)
                    -\alpha^2 \WWW (x_3-x_2))\} +o(R), 
\end{eqnarray}
where $\lim_{R\rightarrow \infty}o(R)/R=0$. 
It is seen  that 
\begin{align}
 \lim_{\alpha\to\infty}\frac{E(k(\alpha)-\alpha^2 \WWW (x_1-x_2)
                    -\alpha^2 \WWW (x_2-x_1))}{\alpha^2}  
 &=[N(N-1)-2] \WWW (0), \\
 \lim_{\alpha\to\infty}\frac{E(k(\alpha)-\alpha^2 \WWW (x_2-x_3)
                    -\alpha^2 \WWW (x_3-x_2))}{\alpha^2}  
 &=[N(N-1)-2] \WWW (0), \\
 \lim_{\alpha\to\infty}\frac{E(k(\alpha))}{\alpha^2} &=N(N-1) \WWW (0). 
\label{y2}
\end{align}
By (W2), we have $\WWW (0)<0$. Therefore combining \kak{y1}-\kak{y2} 
we see that there exists  $\alpha''>0$
such that 
$
\inf\sigma_{\mathrm{ess}}(k(\alpha)) -\is (k(\alpha)) > 0$ for 
$|\alpha|
>\alpha''$. 
This implies the desired result. 
\qed
\end{proof}
\begin{lemma}{\label{delta}}
Let $u_\alpha$ be a normalized ground state of $k(\alpha)$, where $|\alpha|>\alpha''$. 
Then
$  |u_\alpha(y_1,\ldots,y_{N-1})|^2 \to \delta(y_1)\cdots\delta(y_{N-1})$ as $\alpha\to\infty$ in the 
sense of distributions.
\end{lemma}
\begin{proof}
It suffices to show that for all $\epsilon>0$, 
\eq{z6}
  \lim_{\alpha\to\infty} \int_{|Y_0|>\epsilon} |u_\alpha(Y_0)|^2 d Y_0 =0, \quad
 Y_0=(y_1,\ldots,y_{N-1}).
\en
We prove \kak{z6} by a reductive absurdity. Assume that 
$\d  \liminf_{\ell\to\infty} \int_{|Y_0|>\epsilon} |u_{\alpha_\ell}(Y_0)|^2 d Y_0 >0 $ 
for some constant $\epsilon>0$ and some sequence $\{\alpha_\ell\}_{\ell=1}^\infty \subset \mathbb{R}$
such that $\alpha_\ell\to\infty (\ell\to\infty)$. 
We can take a subsequence $\{\hat\alpha_\ell\}_{\ell=1}^\infty \subset \{\alpha_\ell\}_{\ell=1}^\infty$ so that 
$$ \gamma :=\lim_{\ell\to\infty} \int_{|Y_0|>\epsilon} |u_{\hat\alpha_\ell}(Y_0)|^2 d Y_0 >0. $$ 
Since 
$k(\alpha)/\alpha^2  \geq N(N-1)\WWW (0)$ and 
$ \lim_{\alpha\to\infty} E(k(\alpha)/\alpha^2) = N(N-1)\WWW (0)$, 
we have 
\begin{align*}
  N(N-1)\WWW (0) 
  &= \lim_{\ell\to\infty}\frac{1}{\hat\alpha_\ell^2}(u_{\hat\alpha_\ell}, k(\hat\alpha_\ell) u_{\hat\alpha_\ell})=
\lim_{\ell\to\infty} (u_{\hat\alpha_\ell}, \sum_{j\neq l}^N  \WWW (x_j(Y_0)-x_l(Y_0)) u_{\hat\alpha_\ell})  \\
  &  \geq (1-\gamma) N(N-1)W(0) +
     \gamma \inf_{|Y_0|>\epsilon}\sum_{j\neq l}^N   \WWW (x_j(Y_0)-x_l(Y_0)) \\
  &  \geq N(N-1) \WWW (0).
\end{align*}
Thus we have 
\eq{z7}
\inf_{|Y_0|>\epsilon}\sum_{j\neq l}^N   \WWW (x_j(Y_0)-x_l(Y_0)) = N(N-1)W(0).
\en 
By (W2) and \kak{z7} there exists 
a sequence $Z_n=(z_{1,n},\ldots,z_{(N-1),n})\in\mathbb{R}^{d(N-1)}$ such 
that $|Z_n|>\epsilon$ and 
$
 \lim_{n\to\infty}(  x_j(Z_n)-x_l(Z_n)) \to 0$ 
for $j\not =l$. 
By the definition of $x_j(Y)$, we have
\begin{align*}
  &\lim_{n\to\infty}(x_2(Z_n)-x_1(Z_n)) = \lim_{n\to\infty}z_{1,n} =0,  \\
  &\lim_{n\to\infty}(x_3(Z_n)-x_2(Z_n)) = \lim_{n\to\infty}(z_{2,n}-\frac{1}{2}z_{1,n})
                                     = \lim_{n\to\infty}z_{2,n} =0, \\
  & \qquad    \cdots \\
  & \lim_{n\to\infty}(x_N(Z_n)-x_{N-1}(Z_n)) = \lim_{n\to\infty}z_{N-1,n}=0.
\end{align*}
This is a contradiction to $|Z_n|>\epsilon>0$ for all $n$. \qed
\end{proof}

{\it Proof of Theorem \ref{Tex}}\\
Let $u_\alpha$ be a ground state of $k(\alpha)=Uh^0(\alpha) U^{-1}$. 
By Proposition \ref{A2}, 
we may assume that $V \in C_0^\infty(\BR)$. Let $|\alpha|>\alpha''$.   
Let $v\in C_0^\infty(\mathbb{R}^d)$ be a normalized vector such that 
\eq{y3}
( v, (-\frac{1}{2Nm}\Delta_{x_c} + NV(x_c))v ) < 0.
\en 
Such a vector exists by $(W3)$. 
We set 
$ \Psi(Y)=\Psi(x_c,Y_0) := v(x_c)u_\alpha(Y_0)$ for 
$Y=(x_c,Y_0)=(x_c,y_1,\ldots,y_{N-1}) \in \RR^{dN}$. 
Then 
\eq{y4}
  (\Psi, Uh^V(\alpha)U^{-1}\Psi)
  = - \frac{1}{2mN}(  v, \Delta_{x_c} v) 
  + {\cal E}^0(\alpha)
  + ( \Psi, \sum_{j=1}^N V(x_j(Y)) \Psi).
\en 
We define
\begin{align*}
  V_{j,\mathrm{smeared}}^\alpha(x_c) :=
  \int_{\mathbb{R}^{d(N-1)}} d y_1 \cdots d y_{N-1}
  V(x_j(Y)) |u_\alpha(y_1,\ldots,y_{N-1})|^2, \quad j=1,\ldots,N.
\end{align*}
By Lemma \ref{delta}, we have
\begin{align*}
  \lim_{\alpha\to\infty}( \Psi, \sum_{j=1}^N V(x_j(Y)) \Psi)
 =\lim_{\alpha\to\infty}\sum_{j=1}^N ( v, V_{j,\mathrm{smeared}}^\alpha v
)
  = ( v, NV(x_c)v ).
\end{align*}
Therefore, by \kak{y3} and \kak{y4},  
$(\Psi, h^V(\alpha)\Psi) < {\cal E}^0(\alpha)$ 
for $|\alpha|>\alpha'''$ with some $\alpha'''>0$.
By this inequality, Lemma \ref{ene} and \kak{sd}, we conclude that for $\alpha$ with 
$|\alpha|>\alpha_c:=\max\{\alpha',\alpha'''\}$, 
$\Xi^V(\alpha)-{\cal E}^V(\alpha)\geq {\cal E}^0(\alpha)-{\cal E}^V(\alpha)>0$. 
Then the theorem follows. 
\qed
\appendix
\section{The bottom of an essential  spectrum}
We give a general lemma. 
\bl{B1}
Let $K_\epsilon$, $\epsilon>0$,  and $K$ 
 be  self-adjoint operators on a Hilbert space ${\cal K}$ 
and 
$\s_{\rm ess}(K_\epsilon)=[\xi_\epsilon,\infty)$. 
Suppose that $\lim_{\epsilon\rightarrow 0} 
K_\epsilon=K$ in the uniform 
resolvent sense, 
and $\lim_{\epsilon\rightarrow 0} \xi_\epsilon=\xi$. 
Then $\s_{\rm ess}(K)=[\xi,\infty)$. In particular $\lim_{\e\rightarrow 0}\is_{\rm ess}(K_\e)=\is_{\rm ess}(K)$.
\el
\proof 
Let $a>\xi$. Then there 
exists $\e_0$ such that 
for all $\e$ with  $\e<\e_0$, 
$\xi_\e<a$,  
from which we have 
$a\in \s(K_\e)$ for all $\e<\e_0$. 
Since $K_\e$ uniformly converges to $K$ in the resolvent sense, 
$a\in \s(K)$ follows from \cite[Theorem VIII.23 and p.291]{rs1}. 
Since $a$ is arbitrary, 
$(\xi, \infty)\subset \s(K)$ follows and then 
$[\xi,\infty)\subset \s_{\rm ess}(K)$. 
It is enough to show $\is_{\rm ess}(K)=\xi$. 
Let $\lambda \in [\is_{\rm ess}(K),\xi)$ but $\lambda \not\in\s(K)$. 
Note that for all sufficiently small $\e$, $\lambda \not\in\s(K_\e)$ 
by 
\cite[Theorem VIII.24]{rs1}. Since 
$\RR\setminus\s(K)$ is an open set, there exists $\delta>0$ such that 
$(\lambda -\delta, \lambda +\delta)\not\subset \s(K)$. 
Let $P_A(T)$ denote the spectral projection of 
a self-adjoint operator $T$ on a Borel set $A\subset\RR$. 
We have 
$\lim_{\e\rightarrow 0} P_{(\is_{\rm ess}(K)-\delta', \lambda )}(K_\e)
=
P_{(\is_{\rm ess}(K)-\delta', \lambda )}(K)$ uniformly 
by \cite[Theorem VIII.23 (b)]{rs1}. 
In particular, for some $\delta'>0$,   
$$\| P_{(\is_{\rm ess}(K)-\delta', \lambda )}(K_\e)-
P_{(\is_{\rm ess}(K)-\delta', \lambda )}(K)\|<1,$$ 
which implies that 
$P_{(\is_{\rm ess}(K)-\delta', \lambda )}(K_\e){\cal K}$ is isomorphic to 
$P_{(\is_{\rm ess}(K)-\delta', \lambda )}(K){\cal K}$, and then 
 $P_{(\is_{\rm ess}(K)-\delta', \lambda )}(K){\cal K}$ is 
a finite dimensional space, since that of  
$P_{(\is_{\rm ess}(K_\e)-\delta', \lambda )}(K){\cal K}$ is 
finite.
Thus $ (\is_{\rm ess}(K)-\delta', \lambda )\cap \s(K)\subset \s_{\rm disc}(K)$. 
This is a contradiction. Hence we have $[ \is_{\rm ess}(K),\xi)   \subset \s(K)$.
Suppose that $ \is_{\rm ess}(K) <\xi$. 
Let $\tau>0$ be  sufficiently small.
Note that $(\is_{\rm ess}(K)-\tau, \is_{\rm ess}(K)+\tau)\subset \s_{\rm disc}(K_\e)$ 
for all sufficiently small $\e$. 
Let $\theta\in C_0^\infty(\RR)$ satisfy that 
$$\theta(z)=\lkk 
\begin{array}{ll}1,& |z-\is_{\rm ess}(K)|<\tau,\\
0,&|z-\is_{\rm ess}(K)|>2\tau.
\end{array}\right.$$ 
Then we have 
$\lim_{\e\rightarrow 0}\theta(K_\e)=\theta(K)$ uniformly by 
\cite[Theorem VIII.20]{rs1}. 
Since $ \theta(K_\e)$ is 
a finite rank operator for all sufficiently small $\e$, 
$\theta(K)$ has to be a compact operator. 
It contradicts with the fact, however,  
that  the spectrum of $\theta(K)$ is continuous. 
Then we can conclude that $\is_{\rm ess}(K)=\xi$ and 
 the proof is complete.
\qed

Let $V: \BR\to \BR$ be a real measurable function.
\bl{A1}
Let $\Delta$ be the $d$-dimensional Laplacian.
Assume that $V(-\Delta+1)^{-1}$ is  compact. 
Then there exists 
a sequence $\{V(\epsilon)\}_{\epsilon>0}$ such that 
$V(\epsilon)\in   C_0^\infty(\BR)$ and 
$\lim_{\e\rightarrow 0} V(\epsilon)(-\triangle +1)^{-1}=V(-\Delta +1)
$ uniformly.
\el
\proof
Generally, let $A$ be a compact operator and $\{B_n\}_n$  bounded operators such that
 $\slim_{n\to\infty} B_n  = 0$, then $B_n A \to 0$ as $n\to\infty$  in the operator norm.
Since $V(-\Delta+1)^{-1}$ is a compact operator, we obtain that for a sufficiently large $R>0$, 
\eq{z2}
 \| (1-\chi_R) V (-\triangle +1)^{-1} \| < \epsilon/3, \quad
\en 
where $\chi_R$ denotes the characteristic function of $\{x\in\BR| |x|<R\}$.
Let $\chi^{(n)}$ denote the characteristic function of 
$\{ x\in\BR| |V(x)|<n \}$. 
Since 
$(1-\chi^{(n)})\to 0$ strongly as $n\to\infty$, 
\eq{z3}
\| (1-\chi^{(n)})\chi_R V (-\triangle +1)^{-1} \| < \epsilon/3
\en 
for a sufficiently large $n$.
Since $C_0^\infty(\supp(\chi_R\chi^{(n)}))$ is dense in $L^2(\supp(\chi_R\chi^{(n)}))$, 
there exists a sequence  
$\{V_m\}_{m}\subset C_0^\infty(\supp(\chi_R\chi^{(n)}))$ such that
$
 \| V_m - \chi_R\chi^{(n)}V \|_{L^2(\BR)} \to 0$ as $m\to\infty$.
Since $\chi_R\chi^{(n)}V$ has a compact support and is bounded, 
we obtain that 
$\slim_{m\to\infty}V_m= \chi_R\chi^{(n)}V$ as an operator. 
Thus for a sufficiently large $m$, 
\eq{z4}
  \| (V_m -\chi_R\chi^{(n)}V) (-\triangle +1)^{-1} \| < \epsilon/3.
\en 
By \kak{z2}-\kak{z4}  we can obtain that for an arbitrary $\epsilon>0$, 
$ 
\| (V- V_m) (-\triangle + 1)^{-1} \| < \epsilon$
for a sufficiently large $m$.
Thus the lemma follows by setting $V_m=V(\epsilon)$. 
\qed
Let $\beta\subset\N$.  
Set 
$$k^0(\beta):=-\sum_{j\in \beta}\frac{1}{2m_j} \Delta_j
+\sum_{i,j\in\beta} V_{ij},\quad 
k^V(\beta):=h^0(\beta)+\sum_{j\in \beta}V_j$$ with 
$V_i\in L_\mathrm{loc}^2(\BR)$ and  
$V_{ij} \in L_\mathrm{loc}^2(\BR)$ such that 
$V_i (-\triangle +1)^{-1}$ and $V_{ij}(-\triangle +1)^{-1}$ 
are compact operators. 
We define 
$K:=k^V(\N)$. 
Let 
\eq{BB3}
\Xi^V:=\min_{\beta\subsetneqq C_N}\{\is(k^0(\beta))+ \is(k^V(\beta))\}
\en 
be the lowest two cluster threshold of $K$. 
\begin{proposition}{\label{A2}}
There exist sequences 
$\{V_i^\e\}_\epsilon  , \{V_{ij}^\e\}_\epsilon  
\subset  C_0^\infty(\BR)$, $i,j=1,...,N$,  
such that 
$$ (1)\ \lim_{\epsilon\to0} \Xi^V(\epsilon)=\Xi^V,\quad 
(2) \ \lim_{\epsilon\to0} \is_{\rm ess}(K(\e))=\is_{\rm ess}(K),$$
where $\Xi^V(\epsilon)$ (resp. $K(\e)$ ) is 
$\Xi^V$ (resp. $K$)   with $V_i$ and $ V_{ij}$ replaced by
$V_i^\e$ and $V_{ij}^\e$, respectively.
\end{proposition}
\proof 
By Lemma \ref{A1},  there exist  sequences 
$\{V_i^\e\}_{\epsilon>0}  , \{V_{ij}^\e\}_{\epsilon>0}  \subset  C_0^\infty(\BR)$, 
such that 
$$V_i^\e(x_i)(-\Delta_i+1)^{-1}\rightarrow V_i(x_i)(-\Delta_i+1)^{-1}$$ and 
$$V_{ij}^\e(x_i-x_j)(-\Delta_i-\Delta_j+1)^{-1}\rightarrow V_{ij}(x_i-x_j)(-\Delta_i-\Delta_j+1)^{-1}$$
uniformly as $\epsilon\rightarrow 0$ for 
$i,j=1,...,N$.
Hence 
$\is(k^V(\epsilon))$ and $\is(k^0(\epsilon))$ 
converge to $\is(k^V)$ and $\is(k^0)$ 
as $\epsilon\rightarrow 0$, respectively.  
Then (1) 
follows from the definition \kak{BB3}. 
By this and the uniform convergence of $K(\e)$ to $K$ in the resolvent sense, 
Lemma \ref{B1} yields (2). 
\qed


\begin{flushleft}
  {\bf Acknowledgment}
\end{flushleft}
F.H thanks 
Grant-in-Aid  for Science Research (C) 17540181 from JSPS 
for a financial support. 
The second author's work partially supported by Research Fellowship of JSPS 
for Young Scientists.


\end{document}